\documentclass[twocolumn]{aastex631}

\usepackage{color}
\usepackage{epsfig}

\definecolor{darkgreen}{rgb}{0.13, 0.55, 0.13}

\begin{document}

\title{Constraining the Diffusion Coefficient and Cosmic-Ray Acceleration Efficiency using Gamma-ray Emission from the Star-Forming Region RCW~38}

\correspondingauthor{pandey.176@osu.edu}

\author[0009-0003-6803-2420]{Paarmita Pandey} 
\affil{Department of Astronomy, The Ohio State University, 140 W. 18th Ave., Columbus, OH 43210, USA}
\affil{Center for Cosmology and Astroparticle Physics, The Ohio State University, 191 W. Woodruff Ave., Columbus, OH 43210, USA}

\author[0000-0002-1790-3148]{Laura A.~Lopez}
\affil{Department of Astronomy, The Ohio State University, 140 W. 18th Ave., Columbus, OH 43210, USA}
\affil{Center for Cosmology and Astroparticle Physics, The Ohio State University, 191 W. Woodruff Ave., Columbus, OH 43210, USA}

\author[0000-0003-4423-0660]{Anna L.~Rosen}
\affil{Department of Astronomy, San Diego State University, San Diego, CA 92182, USA}
\affil{Computational Science Research Center, San Diego State University, San Diego, CA 92182, USA}

\author[0000-0003-2377-9574]{Todd A.~Thompson}
\affil{Department of Astronomy, The Ohio State University, 140 W. 18th Ave., Columbus, OH 43210, USA}
\affil{Center for Cosmology and Astroparticle Physics, The Ohio State University, 191 W. Woodruff Ave., Columbus, OH 43210, USA}

\author[0000-0001-9888-0971]{Tim Linden}
\affil{Stockholm University and The Oskar Klein Centre for Cosmoparticle Physics, Alba Nova, 10691 Stockholm, Sweden}

\author[0000-0002-7329-560X]{Ian Blackstone}
\affil{Department of Physics, Ohio State University, 191 W. Woodruff Ave, Columbus, OH 43210}
\affil{Center for Cosmology and Astroparticle Physics, The Ohio State University, 191 W. Woodruff Ave., Columbus, OH 43210, USA}

\begin{abstract}

Stellar winds from massive stars may be significant sources of cosmic rays (CRs). To investigate this connection, we report a detailed study of gamma-ray emission near the young Milky Way star cluster ($\approx$ 0.5~Myr old) in the star-forming region RCW 38 and compare this emission to its stellar wind properties and diffuse X-ray emission. Using 15 years of Fermi-LAT data in the 0.2 $-$ 300 GeV band, we find a significant ($ \sigma > 22$) detection coincident with the star cluster, producing a total $\gamma$-ray luminosity (extrapolated over 0.1 $-$ 500 GeV) of $L_{\gamma} = (2.66\pm 0.92) \times 10^{34}$ erg s$^{-1}$ adopting a power-law spectral model ($\Gamma = 2.34\pm0.04$). Using an empirical relationship and Starburst99, we estimate the total wind power to be $8 \times 10^{36}$ erg s$^{-1}$, corresponding to a CR acceleration efficiency of $\eta_{\rm CR} \simeq 0.4$ for an assumed diffusion coefficient consistent with $D = 10^{28}$~cm$^{2}$~s$^{-1}$. Alternatively, a lower acceleration efficiency of 0.1 can produce this $L_{\gamma}$ if the diffusion coefficient is smaller, $D\simeq 2.5\times10^{27}\,{\rm cm^2\,\,s^{-1}}$. Additionally, we analyze Chandra X-ray data from the region and compare the hot-gas pressure to the CR pressure. We find the former is four orders of magnitude greater, suggesting that the CR pressure is not dynamically important relative to stellar winds. As RCW~38 is too young for supernovae to have occurred, the high CR acceleration efficiency in RCW~38 demonstrates that stellar winds may be an important source of Galactic CRs. 
\end{abstract}

\keywords{Galactic cosmic rays (567), Gamma-ray astronomy (628), Young star clusters (1833), Stellar winds (1636), Stellar feedback (1602), X-ray sources (1822)}

\section{Introduction} \label{sec:intro}

Cosmic rays (CR) are relativistic charged particles that are a fundamental component of the Galaxy. CRs have a direct influence on the thermodynamics and chemistry of the interstellar and circumgalactic medium \citep{Boulares, Zweibel, Padovani20}. They contribute to the ionization and heating in molecular clouds where UV and X-ray photons are shielded \citep{Dalgarno}, and CR spallation is the primary mechanism of production of light elements such as Li, Be, and B \citep{Field1, Field2, Ramaty}. On galactic scales, the role of CRs in galaxy formation has received increased attention in recent years \citep{Ruszkowski17, Chan19, PFHopkins20}. In particular, galaxy formation simulations show that CR feedback may play a vital role in the launching of galactic-scale winds \citep{Booth, Salem, Pakmor, Simpson, Jacob, Modak} and alter the phase structure of the circumgalactic medium (e.g., \citealt{Salem16,Butsky23}). 

In order to model CR feedback in galaxies, it is crucial to constrain CR acceleration and transport observationally. CR electrons are detectable at radio and X-ray wavelengths, but it is critical to probe CR protons as they comprise the bulk of the CR energy. Fortunately, emission associated with CR protons is observable with $\gamma$-ray facilities: when CR protons collide with gas, neutral pions are produced that decay and dominate the GeV emission of star-forming galaxies (e.g., in the Milky Way: \citealt{Strong10}).

The origin of galactic CRs is a subject of ongoing discussion. Diffusive shock acceleration (DSA) in supernova remnants (SNRs) has long been considered the primary source of galactic CRs \citep{Zwicky, Drury, Blasi}. However, there is increasing evidence showing that stellar wind feedback from young massive star clusters (YMCs) may also contribute to the high-energy CR budget of star-forming galaxies \citep{Aharoniannature}. First recognized as possible CR accelerators in the 1980s \citep{Casse80,Cesarsky83}, particles may be accelerated either in the immediate vicinity of the stars through the shocks produced by collisions of individual stellar winds,
through the reverse shock of collective winds interacting with the surrounding medium, or the forward shock at the front of the expansion in superbubbles \citep{Parizot04,Bykov14, Gupta2, Gupta1}. 

Since the advent of modern GeV and TeV facilities, several star clusters have now been detected in $\gamma$-rays: e.g., the Cygnus cocoon \citep{Cygnus,Aharoniannature,Cyg2},  Westerlund 2 \citep{Westerlund2}, NGC 3603 \citep{NGC36031} and \citep{NGC36032}, M17 \citep{Liu22}, W40 \citep{W40}, and W43 \citep{W43}. There are many more known YMCs in the Milky Way (see \citealt{zwart}), a majority of which do not have reported $\gamma$-ray detections yet. One reason is that most of these sources are located within the Galactic plane, and it is challenging to confirm $\gamma$-ray associations in crowded regions due to the limited spatial resolution at GeV energies. YMCs with ages $<$3~Myr without nearby SNRs or pulsars are especially valuable targets to evaluate the efficacy of stellar winds as CR accelerators because these YMCs are too young for supernovae to have occurred. 

In this paper, we present the Fermi Gamma-ray Space Telescope detection of the star-forming region RCW~38. Positioned one degree south of the Galactic plane and with an estimated age of 0.1 $-$ 0.5~Myr \citep{Wolk06, Fukui16}, it is an optimal target to search for $\gamma$-rays and constrain the efficiency of CR acceleration from the collective stellar winds in a YMC.

RCW~38 is located 1.7 kpc away and is powered by an embedded (extinction coefficient $A_{\rm V} \sim10$ mag) star cluster \citep{Wolk06}. There are two defining IR sources in RCW~38; the brightest source at 2 $\mu$m is IRS~2 corresponds to an O5.5 binary located at its center \citep{DeRose09}. The brightest source at 10 $\mu$m is IRS~1, a dust ridge that extends 0.1 $-$ 0.2 pc in the north-south direction \citep{Wolk06, Kuhn_2015b, Kuhn_2015a}. CO observations found a total cloud mass of $2.3 \times 10^4 \ M_\odot$ \citep{Fukui16}  and that the star cluster likely formed via a cloud-cloud collision, making it the third identified YMC in the Milky Way formed this way (the others being NGC~3603 \citep{Fukui2014} and Westerlund~2 \citep{Fukui2009, Ohama2010}). \cite{Muzic} performed an extensive study of the low-mass stellar content in RCW~38's star cluster using NACO/VLT data and found it has a top-heavy initial mass function (IMF) that is shallower than a Salpeter IMF \citep{Salpeter} or a Kroupa IMF \citep{Kroupa}, with $ \rm dN/dM \propto  M^{-\alpha}$, where $\alpha = 1.60 \pm 0.13$. Within the central few parsecs, the region harbors $\sim10^{4}$ stars, 20 of which are confirmed O-type stars (and nearly 30 total candidates) \citep{Wolk06, Broos, Kuhn_2015a}.
\begin{figure}
    \begin{center}       
\includegraphics[width=\columnwidth]{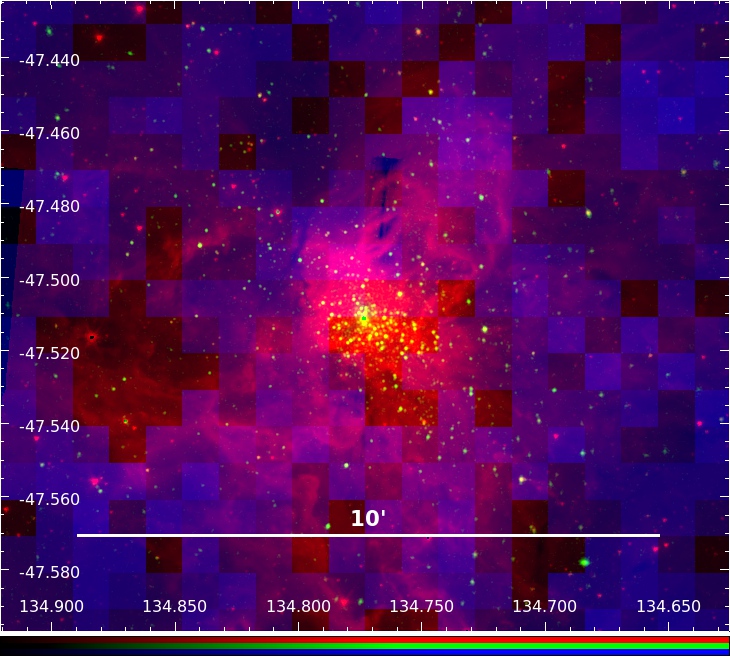}
       \caption{Multiwavelength image of RCW 38, with 3.6$\mu$m IR Spitzer image in red \citep{Wolk06}, the Chandra X-ray broad-band ($0.5-7.0~$keV) in green, and the background subtracted $2-300$ GeV Fermi-LAT counts map in blue. The $\gamma$-ray emission region is coincident with the IR and X-ray emission.} 
       \label{fig:multi}
    \end{center}    
\end{figure}

The Fermi-LAT data above 1~GeV from RCW~38 has been previously studied by \cite{Peron24} and \cite{Ge2024}. Both papers suggest that the observed GeV emission is caused by the interaction of CR protons accelerated by the stellar winds with the ambient gas. This paper considers the Fermi-LAT data from 0.2 $-$ 300 GeV as well, expanding upon the previous work of by estimating the wind luminosity from observations (rather than adopting a fiducial value) and taking advantage of the Chandra X-ray data toward RCW~38 to explore the relationship between the hot gas and CRs. Specifically, we use an empirical relation between bolometric luminosity, mass-loss rate, and stellar wind velocity of the stars in RCW 38 to estimate the fraction of the energy injected by stellar winds is responsible for accelerating CRs. We then compare our estimate with that of simulated results by Starburst99. Our work focuses primarily on the observational signatures of stellar wind feedback in young clusters with an emphasis on connecting current theoretical predictions to observational analysis.

This paper is organized as follows. We present the Fermi $\gamma$-ray analysis in Section~\ref{sec:fermianalysis}, including spatial and likelihood analysis confirming the association with RCW~38 (Section~\ref{sec:spatial} and Section~\ref{sec:likelihood}). We find evidence that the $\gamma$-ray emission is extended in Section~\ref{sec:extension}, and we produce the $\gamma$-ray spectral energy distribution (SED) and estimate the $\gamma$-ray luminosity in Section~\ref{sec:spectral_analysis}. In Section~\ref{sec:Xrayanalysis}, we analyze archival Chandra X-ray data toward RCW~38 in order to measure the hot-gas properties produced by the stellar winds in the region. In Section~\ref{sec:wind_power}, we evaluate the bolometric luminosity and wind power of the region. In Section~\ref{sec:CRinjection}, we argue that if the gamma-ray emission is hadronic and if the CR losses are dominated by diffusion, then the observations necessitate either a high CR acceleration efficiency or a relatively small diffusion coefficient relative to typical ISM values. In Section~\ref{sec:CR_pressure}, we find that the CR pressure is likely much weaker than the hot gas pressure, indicating that CR pressure is not dynamically important on the size scales of YMCs and their surrounding HII regions. In Section~\ref{sec:leptons}, we show that CR leptons are not likely to contribute to the detected $\gamma$-ray flux thereby confirming the observed $\gamma$-ray emission originates from the destruction of CR protons. In Section~\ref{sec:Comparison}, we compare and contrast our work with that previously conducted by \citet{Peron24} and \cite{Ge2024} on RCW~38. In Section~\ref{sec:conclusions}, we summarize our conclusions.  

\section{Data Analysis and Results}

\subsection{Fermi-LAT Data Analysis} \label{sec:fermianalysis}

We used data from the Fermi $\gamma$-ray Space Telescope, which was launched in 2008 and houses two scientific instruments, the Large Area Telescope (LAT) and the Gamma-ray Burst Monitor (GBM). LAT onboard Fermi observes $\gamma$-ray photons utilizing electron-positron pair production in a silicon tracker and detects $\gamma$-rays in the energy range $0.1–300$~GeV. It has a spatial resolution of $<1^{\circ}$ for E$>$1~GeV, a very wide field of view ($\sim$2.4~sr), and an effective area $>$8000~cm$^2$ \citep{LAT}. The latest upgrade to the event reconstruction process and instrumental response functions (referred to as Pass 8) improved the effective area, the accuracy of point spread function, and the system's ability to reject cosmic-ray backgrounds \citep{pass8}.

\subsubsection{Spatial Analysis} 
\label{sec:spatial}

In our analysis, we utilized nearly 15 years of data spanning from August 8, 2008  (MET 239846401) to June 5, 2023 (MET 707616005) of LAT events in the reconstructed energy range from 200 MeV to 300 GeV within a $20^{\circ}$ region of interest (ROI) at the optical coordinates of RCW~38 (R.A. $=134.773^{\circ}$, Dec $=-47.5109^{\circ}$). We analyzed the data using Fermitools analysis package (v11r5p3\footnote{\href{https://github.com/fermi-lat/Fermitools-conda}{https://github.com/fermi-lat/Fermitools-conda}}), and we evaluated the extension and spectral energy distribution of the source using the FermiPy python package \citep{Wood_fermipy}. 

We used {\it gtselect} to select photons of energies 200 MeV $-$ 300 GeV with an arrival direction $<90^{\circ}$ from the local zenith to remove contamination from $\gamma$-rays produced by CR interactions in the upper layers of Earth’s atmosphere. The good time intervals (GTIs) when the telescope was operating normally were selected using the filters “DATA QUAL $>$ 0” and “LAT CONFIG==1”. The {\fontfamily{cmtt}\selectfont P8R3\_SOURCE\_V3} instrument response was used for analysis.

\begin{figure}
    \begin{center}       
\includegraphics[width=\columnwidth]{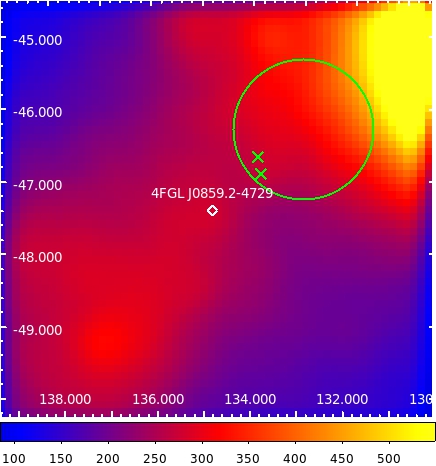}
       \caption{Counts map of the  $6^{\circ}\times6^{\circ}$ region centered on 4FGL J0859.2-4729 in the 0.2 $-$ 300~GeV band. The green circle shows the extension of the Vela Jr supernova remnant. The two Xs show the position of PSR J0855-4644 and PSR J0855-4658 which are present within $1^{\circ}$ of J0859.2-4729.} 
       \label{fig:countsmap}
    \end{center}    
\end{figure}

We performed a binned maximum-likelihood analysis to estimate the best-fit model parameters using a $20^{\circ}$ square region centered on RCW~38 with ten equally spaced logarithmic bins in energy. We selected events only belonging to the SOURCE class (evclass=128) and evtype=3 (corresponding to standard analysis in Pass 8) within the ROI. We incorporated the Galactic and extragalactic diffuse emission and isotropic background emission in the model via the templates {\fontfamily{cmtt}\selectfont gll\_iem\_v07} and {\fontfamily{cmtt}\selectfont iso\_P8R3\_SOURCE\_V6\_v06} (tilt fit has not been considered). The $\gamma$-ray data were modeled with the comprehensive Fermi-LAT source catalog, 4FGL-DR3 \citep{DR3}. 

Figure~\ref{fig:multi} shows the $2-300$~GeV background subtracted Fermi count map in blue (produced using {\it gtbin}) in the vicinity of RCW~38 compared to the Spitzer 3.6$\mu$m in red \citep{Winston12} and the Chandra broad-band ($0.5-7.0$~keV) X-rays in green. The $68\%$ confidence interval of the spatial resolution of the LAT is $\approx0.6^{\circ}$ at 2 GeV and $\approx0.1^{\circ}$ at 100 GeV, compared to a resolution of $\sim3^{\circ}$ at 200 MeV\footnote{slac.stanford.edu/exp/glast/groups/canda/lat\_Performance.htm}. Thus, we consider the $2-300$~GeV emission for spatial analysis and multiwavelength comparison, and we find the gamma-rays are spatially coincident with the point sources and diffuse emission detected in the IR and X-rays.

\subsubsection{Likelihood analysis}
\label{sec:likelihood}

We employed the maximum likelihood technique to investigate the $\gamma$-ray emission quantitatively. The Fermitools command \textit{gtlike} computes the best-fit parameters by maximizing the joint probability of getting the observed data from the input model, given a specified model for the distribution of gamma-ray sources on the sky and their spectra. The likelihood $\mathcal{L}$ is the likelihood that our spatial and spectral model accurately captures the data. The test statistic (TS) is defined as $\rm TS  = -2\,\rm \ln(\mathcal{L}_{0}/\mathcal{L}_{1}$), where $\mathcal{L}_{0}$ and $\mathcal{L}_{1}$ are the likelihoods without and with the addition of a source at a given position, respectively.

We used {\it gtlike} to run a binned likelihood analysis over the energy range of 200 MeV $-$ 300 GeV. The spectral indices and normalizations of the sources within $6^{\circ}$ of the source, together with the normalization of the Galactic diffuse emission and isotropic component, were free parameters in the fit. Any sources located beyond a radius of 6$^{\circ}$ from the target and those with significance less than $5\sigma$ were fixed to the spectral parameter values given in 4FGL-DR3\footnote{\href{https://fermi.gsfc.nasa.gov/ssc/data/analysis/user/readme_make4FGLxml.txt}{readme\_make4FGLxml.txt}}.

The source 4FGL J0859.2$-$4729 is located only 0.03$^{\circ}$ away from the optical coordinates of RCW~38, and thus it is a possible $\gamma$-ray counterpart to the star-forming region. From the 4FGL catalog, this source is not variable, and its initial TS is 495 using a log-parabola (LP) spectral model in 4FGL-DR3. There are $\approx800$ Fermi 4FGL sources within $\pm 2^{\circ}$ of the galactic plane \citep{Ballet2023}. The probability that a random source is within 0.03$^{\circ}$ of the star cluster is  $\frac{0.03^2}{1500}  \simeq 6\times 10^{-7}$. Consequently, we get the probability that one of the sources is coincident is $\simeq 5\times10^{-4}$. Thus, we interpret that this 4FGL source is likely associated with RCW~38. 

According to the Australia Telescope National Facility (ATNF) Pulsar Catalog \citep{ATNF2005}, two pulsars, PSR J0855$-$4644 and PSR J0855$-$4658, are located within $1^{\circ}$ of RCW~38. Fermi-LAT pulsars are frequently classified as exhibiting either a power-law or a power-law with a spectral cutoff spectrum, usually having a cutoff at energies below 10 GeV \citep{AbdoPulsar2013}. Although we cannot discard the possibility that the observed $\gamma$-ray emission originates from the pulsars, it appears improbable, given that a considerable amount of the emission is recorded $>$ 10 GeV range (see Fig \ref{fig:sed}). \cite{Ge2024} state that PSR J0855$-$4658 could not contribute $\gamma$-ray emission due to its low spin-down power, and PSR J0855$-$4658's location $0.8^{\circ}$ away from the $\gamma$-ray peak make it unlikely as the only source as well. It is plausible that these pulsars may contribute non-negligibly to the total $\gamma$-ray emission. Although the Vela Jr supernova remnant and the Vela X pulsar are bright and in the ROI, they are $\sim$1.68$^{\circ}$ and $\sim$4.61$^{\circ}$ away, respectively, sufficiently far for the Fermi PSF to resolve them. Figure \ref{fig:countsmap} highlights the spatial distribution of $\gamma$-ray emission in the $0.2-300$ energy band including a $6^{\circ}\times6^{\circ}$ region centered on 4FGL J0859.2$-$4729. The two pulsars are denoted by X, whereas the green circle denotes the Vela Jr supernova remnant. Both are spatially offset from the peak of the 4FGL source's $\gamma$-ray emission.

To investigate the association of the 4FGL source with RCW~38 and the inherent distribution of accelerated particles, we modeled the $\gamma$-ray spectrum of 4FGL J0859.2$-$4729. In our maximum likelihood analysis, we tested a power-law (PL) and a LP spectral model. The PL model is defined as
\begin{equation}   
    \frac{dN(E)}{dE} = N_0 \left(\frac{E}{E_{\rm p}}\right)^{-\Gamma},
\end{equation}   
where $\Gamma$ is the spectral index, $N_0$ is the pre-factor index (with units of ph~cm$^{-2}$~$\rm{s}^{-1}$~MeV$^{-1}$), and $E_{\rm p}$ is the pivot energy, which is the energy at which error on differential flux is minimal. The LP model is defined as
\begin{equation}
   \frac{dN(E)}{dE} = N_0 \left( \frac{E}{E_p}\right)^{-(\alpha + \beta {\rm log} \frac{E}{E_p})},
\end{equation}
where $N_0$ is the pre-factor index (with units of ph $\rm{cm}^{-2}$~$\rm{s}^{-1}$~$\rm{MeV}^{-1}$), and $\alpha$ and $\beta$ are the spectral index and curvature parameter, respectively. 

\begin{deluxetable*}{lclr}
\label{tab:table1}
\tablenum{1}
\tablecaption{List of different models used for spectral and spatial analysis and their corresponding Log-likelihood values. 
\label{tab:messer}}
\tablewidth{0pt}
\tablehead{
\colhead{Source Position} &   
\colhead{Spectral Model} &
\colhead{Source Type} &
\colhead{$\Delta$TS\tablenotemark{b}} }
\startdata
4FGL~J0859.2-4729\tablenotemark{a} & Log-Parabola & Point source & --  \\
4FGL~J0859.2-4729 & Power-law & Point source & 15.1 \\
Optical coordinates & Log-Parabola & Point source & $-$2.3 \\
Optical coordinates & Power-law & Point source & 11.0 \\
4FGL~J0859.2-4729 & Power-Law & Radial Disk & 27\tablenotemark{c}  \\
4FGL~J0859.2-4729 & Power-law & Radial Gaussian & 35\tablenotemark{c} \\
\enddata
\tablenotetext{a}{This model is from the 4FGL-DR3 catalog, yielding a TS = 495 and a log-likelihood of $-$16624147.3.}
\tablenotetext{b}{$\Delta$TS gives the improvement in log-likelihood relative to the model in the 4FGL-DR3 catalog.}
\tablenotetext{c}{$\Delta$TS gives the improvement in log-likelihood for the best-fit model for extension relative to the no-extension (point-source) scenario.}
\end{deluxetable*}

The results of the likelihood analysis for the two different models of 4FGL J0859.2$-$4729 (i.e., adopting the 4FGL-DR3 position) are listed in Table~\ref{tab:table1}. Relative to the 4FGL-DR3 TS value of 495 for a LP model, a PL gives TS of 509 corresponding to a $\sigma > 22$ significance detection in the 200 MeV $-$ 300 GeV band. We obtain a $\Delta$TS = 15.1 for the PL model, where $\Delta$TS is the improvement in log-likelihood relative to the model
in the 4FGL-DR3 catalog. By comparison, the LP and PL models located at the optical coordinates of RCW~38 had $\Delta$TS = $-$2.4 and $\Delta$TS = 12. Given that a PL model has a higher likelihood value for 4FGL J0859.2$-$4729 and carries one less degree of freedom with respect to the LP model, we conclude that the best characterization for the $\gamma$-ray emission that is coincident with RCW~38 is a PL spectrum.

To localize the $\gamma$-ray emission further, we produced a TS map of the $2 - 300$ GeV emission using the command {\it gttsmap}, which computes the improvement in likelihood if a point source is added to each spatial bin. To produce this TS map, we adopted the best-fit model output by {\it gtlike}, removed the source associated with RCW~38, and then computed the TS value for 0.05$^{\circ}$ pixels. 

\begin{figure}
    \begin{center}       
\includegraphics[width=\columnwidth]{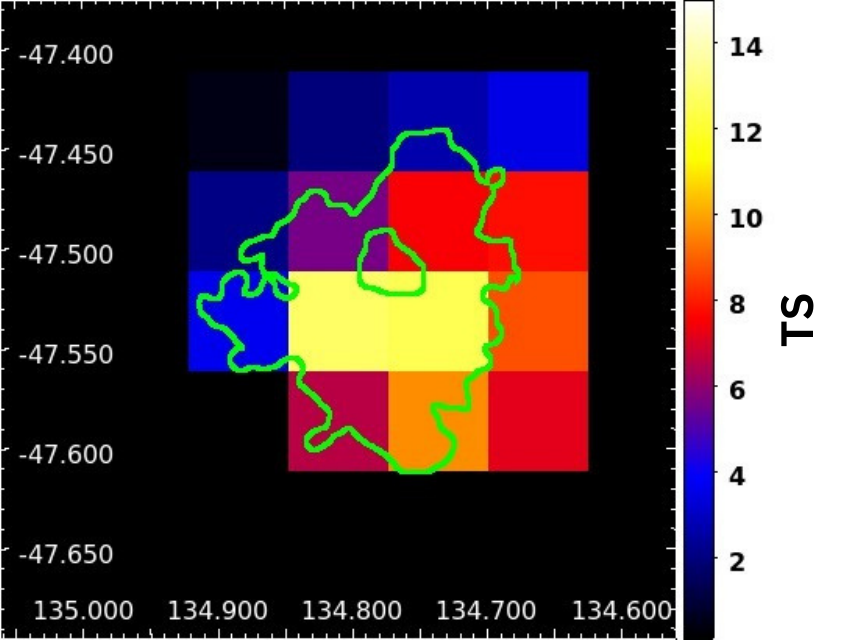}
       \caption{TS map of the 2 $-$ 300~GeV band centered on RCW 38 with a pixel size of $0.05^{\circ}\times0.05^{\circ}$. The green contours reflect the distribution of the 3.6$\mu$m emission of the star-forming region observed by Spitzer \citep{Winston12}, with the inner contour showing the extent of the star cluster powering RCW~38. The maximum TS value of $\approx$13 in the central two pixels is spatially coincident with the star cluster and corresponds to a $\approx3.6\sigma$ detection in the 2 $-$ 300 GeV band.} 
       \label{fig:tsmap}
    \end{center}    
\end{figure}

Figure~\ref{fig:tsmap} gives the TS map of RCW 38 in the 2 $-$ 300 GeV band, with green contours reflecting the distribution of 3.6$\mu$m emission observed by Spitzer \citep{Winston12}. The greatest TS value of $\approx$13 in the central two pixels is coincident with the star cluster powering RCW~38 and corresponds to a 3.6$\sigma$ detection in the 2 $-$ 300 GeV band. 

\subsubsection{Extension Analysis}
\label{sec:extension}

To investigate whether 4FGL J0859.2$-$4729 is a point source or is extended, we conducted extension tests in FermiPy utilizing the {\fontfamily{cmtt}\selectfont GTAnalysis.extension} method. We also perform a simultaneous fit to the source position. We explored two spatial models, $Radial \ disk$ and a $Radial \ Gaussian$, that have symmetric, two-dimensional shapes, where the radius $R$ and the width $\sigma$ parameters dictate the size of the source. In both cases, we adopted the PL spectral model. When determining the optimal spatial extension for both templates, we let the galactic and isotropic backgrounds be free parameters. We also let the normalizations of sources within $6^{\circ}$ of the target be free. We found that the best fit corresponds to a spatial template of a Radial Gaussian with extension size of $\sigma = 0.24^{\circ} \pm 0.04^{\circ}$ for 4FGL J0859.2$-$4729 with a $\Delta$TS = 35 ($\approx 6 \sigma$ improvement) relative to the point source spatial model. The new best-fit position for 4FGL J0859.2$-$4729 is R.A. $=134.804^{\circ}$, Dec $=-47.479^{\circ}$ as compared to the previous position of R.A. $=134.804^{\circ}$, Dec $=-47.488^{\circ}$.

\subsubsection{Spectral Analysis} \label{sec:spectral_analysis}

\begin{figure*}
    \begin{center}
      \includegraphics[width=0.9\textwidth]{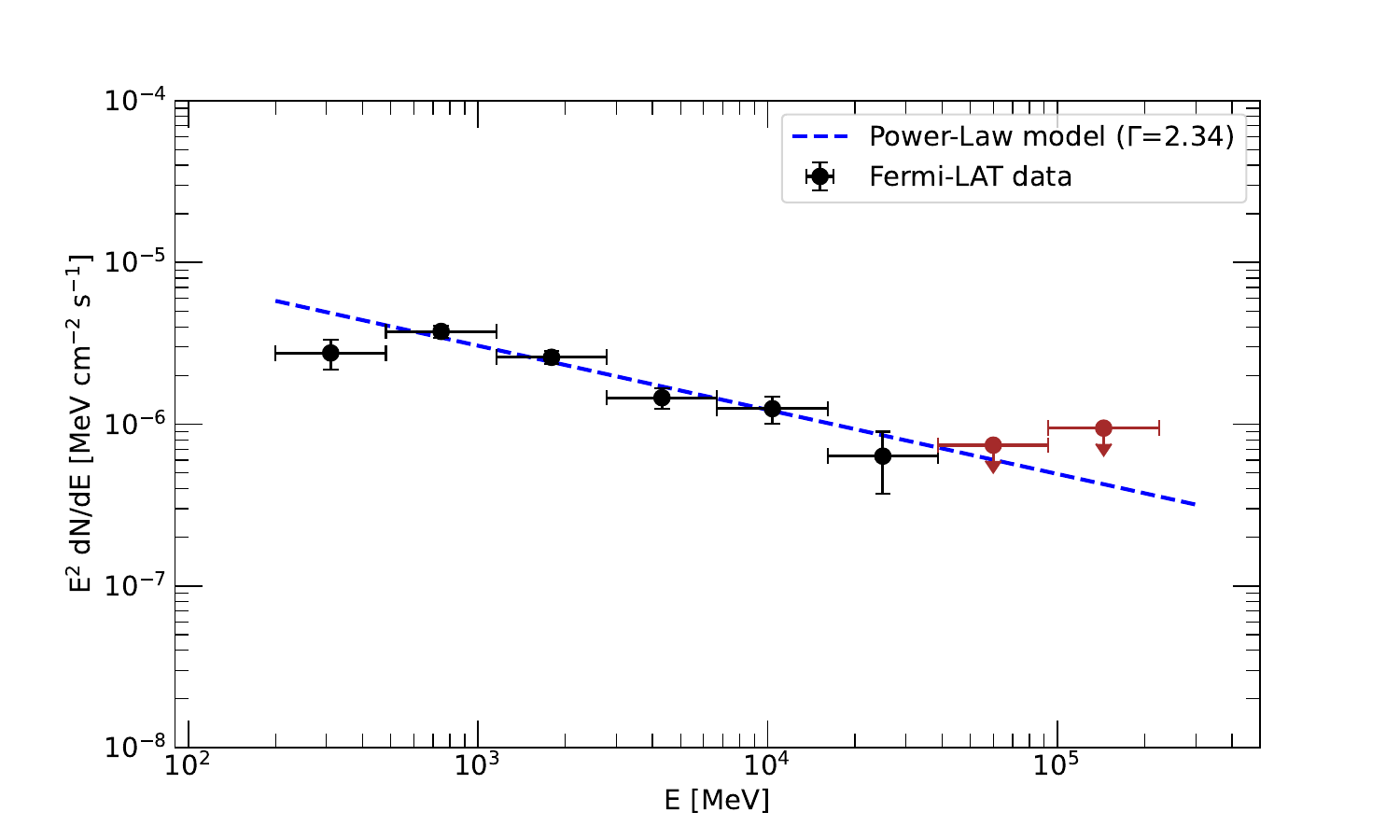}
        \caption{Fermi $\gamma$-ray SED of RCW 38. For each data point, the error bar reflects the statistical uncertainty caused by the effective area. The blue dashed line represents the best-fit PL model which has a photon index of $\Gamma = 2.34\pm 0.04$. The total photon flux produced by the best-fit model in the 0.1 $-$ 500 GeV band is $\Phi^{> 100 \rm MeV}_{\gamma} = (2.35\pm0.17) \times 10^{-8}$ ph cm$^{-2}$ s$^{-1}$, corresponding to a gamma-ray luminosity of $L_\gamma = (2.66\pm 0.92)\times 10^{34}$ erg s$^{-1}$ for a distance of 1.7~kpc.}  
       \label{fig:sed}
    \end{center}  
    \label{pos}
\end{figure*}

To produce the spectral energy distribution (SED) of the $\gamma$-ray emission of 4FGL J0859.2$-$4729 using the PL spectral model and Radial Gaussian spatial distribution (see Section~\ref{sec:extension}), we consider data from the energy range of $0.2 - 300$ GeV, which was then used to extrapolate the $0.1 - 500$ GeV total $\gamma$-ray luminosity. We estimated the SED by varying
the normalization of the PL model independently in 8 energy
bins spaced uniformly in log space from $0.2 - 300$ GeV. Figure~\ref{fig:sed} shows the integrated gamma-ray spectrum with the errors plotted for each data point and the best-fit PL model overplotted. Photons were not detected in the two highest-energy bins above 40 GeV, so the 2-$\sigma$ upper limit for those bins is given.

Figure \ref{fig:sed} shows the integrated gamma-ray spectrum with the errors plotted for each data point and the best-fit PL model overplotted. Photons were not detected in the two highest-energy bins above 40 GeV, so a 2-$\sigma$ upper limit for that bin is given. After performing the spectral analysis, we find a best-fit photon index $\Gamma$ in the PL spectral model to be $\Gamma = 2.34\pm 0.04$, producing a total photon flux in the $0.1-500$ GeV range of $\Phi^{> 100 \rm MeV}_{\gamma} = (2.35\pm0.17) \times 10^{-8}$ ph cm$^{-2}$ s$^{-1}$. Assuming a distance of 1.7 kpc to RCW 38, the associated energy flux corresponds to a luminosity of $L_\gamma = (2.66\pm 0.92)\times 10^{34}$ erg~s$^{-1}$.  
\subsection{Chandra X-ray Data Analysis}
\label{sec:Xrayanalysis}

To determine the properties of the hot ($\sim10^{7}$~K), diffuse gas produced by the shock-heating from stellar winds, we analyzed archival Chandra X-ray observations of RCW~38. Specifically, we used these data to check the spatial extension of RCW~38 and to compute the temperature $kT$, electron density $n_{\rm e}$, pressure $P_{\rm X}$, and X-ray luminosity ($L_{\rm X}$) based on the thermal bremsstrahlung continuum emission. These values are employed in Section~\ref{sec:discussion} to derive the effective number density of nucleons $n_{\rm eff}$ and to compare to the CR pressure $P_{\rm CR}$ in Section~\ref{sec:CR_pressure}.

RCW~38 was observed by Chandra four times totaling 190~ks with the ACIS-I array: for 97~ks in December 2001 (ObsID 2556), for 15 and 40~ks in June 2015 (ObsIDs 16657 and 17681, respectively), and for 38~ks in August 2015 (ObsID 17554). These data were downloaded from the Chandra archive and reduced using the Chandra Interactive Analysis of Observations {\sc ciao} version 4.14 \citep{CIAO2006}. Data were reprocessed (using the {\it repro} command), combined, and exposure corrected (using the {\it merge\_obs} function). 

\begin{figure}[t]
    \begin{center}
      \includegraphics[width=\columnwidth]{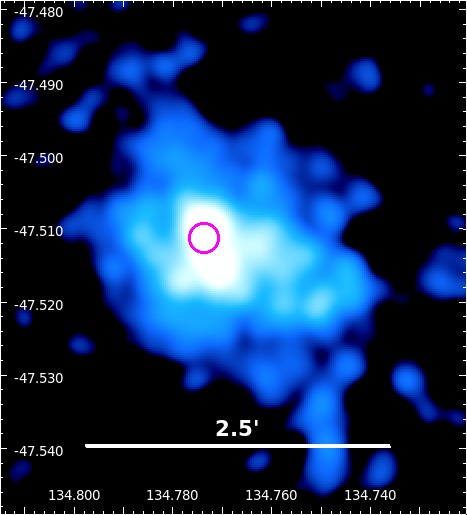}
       \caption{Hard X-rays ($2-7$ keV) from RCW 38 with point sources removed. The IRS 2 binary star system is highlighted by the magenta circle.} 
       \label{fig:hardXrays}
    \end{center}
    \label{pos}
\end{figure}

Figure~\ref{fig:multi} shows the exposure-corrected, broad-band (0.5$-$7.0 keV) Chandra image in green, with hundreds of apparent point sources (as identified by \citealt{Wolk06}). To map the diffuse X-ray emission and measure its extent, point sources were identified (using the {\it wavdetect} function) and removed (using the {\it dmfilth} command). Figure~\ref{fig:hardXrays} shows the resulting diffuse hard X-ray ($2.0-7.0$~keV) map that has an angular extent of $\approx$2.5\arcmin\ and is dominated by non-thermal emission (based on our spectral fits described below and results from \citealt{Wolk02} and \citealt{Wolk06}). The centrally enhanced region corresponds to the location of the O5.5 binary in IRS~2 \citep{DeRose09}. 

\begin{figure}[t]
    \begin{center}
\includegraphics[width=\columnwidth]{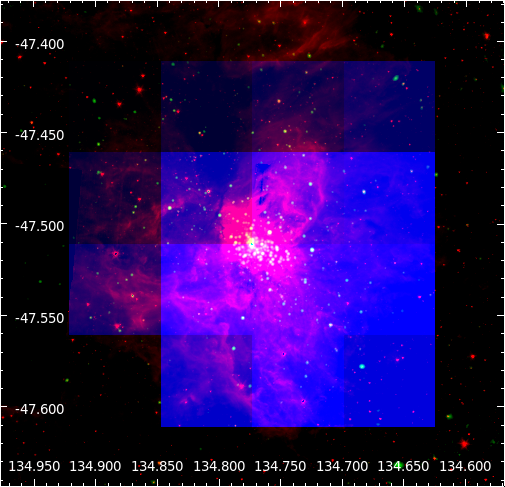}
       \caption{Three color image of RCW 38 combining Chandra broad-band X-ray data (green), Fermi $\gamma$-ray TS map of 1 degree (blue), and Spitzer infrared data (red). The $\gamma$-ray TS map represents the energy range of $2-300$~GeV with $0.05-$degree spatial resolution.} 
       \label{fig:threecolor}
    \end{center}
     \end{figure}

Figure~\ref{fig:threecolor} compares the Fermi $\gamma$-ray, Chandra X-ray, and Spitzer (3.6$\mu$m) IR images of RCW~38. As noted above, the $\gamma$-ray emission is coincident with the star cluster seen in X-rays and the larger star-forming complex traced by the IR. 

To estimate the hot gas properties of RCW~38, we extracted and modeled source X-ray spectra from a 2\arcmin\ radius circular region, with interior point sources excluded. We subtracted background spectra from a circular region $\approx$ 6\arcmin\ southwest of RCW~38 that was 0.5\arcmin\ in radius. The background-subtracted source spectra from each observation were modeled simultaneously using XSPEC Version 12.12 \citep{XSPEC}. The model included a multiplicative constant component (\textsc{const}), one absorption (\textsc{phabs}) component, a power-law component (\textsc{powerlaw}), and one optically thin, thermal plasma component (\textsc{apec}). The \textsc{const} component was allowed to vary and accounted for slight variations in emission between the observations. The \textsc{phabs} component accounted for the galactic absorption in the direction of RCW~38 and was allowed to vary. The \textsc{powerlaw} component accounted for the non-thermal X-ray emission, and the \textsc{apec} component represented the thermal plasma. We found that both a thermal and power-law component were necessary to adequately fit the spectra, consistent with the results of \cite{Wolk02} and \cite{Wolk06}. We fixed abundances to solar values from \cite{Asplund2009} and photoionization cross sections from \cite{Verner1996}.

\begin{figure}[t]
    \begin{center}
      \includegraphics[width=\columnwidth]{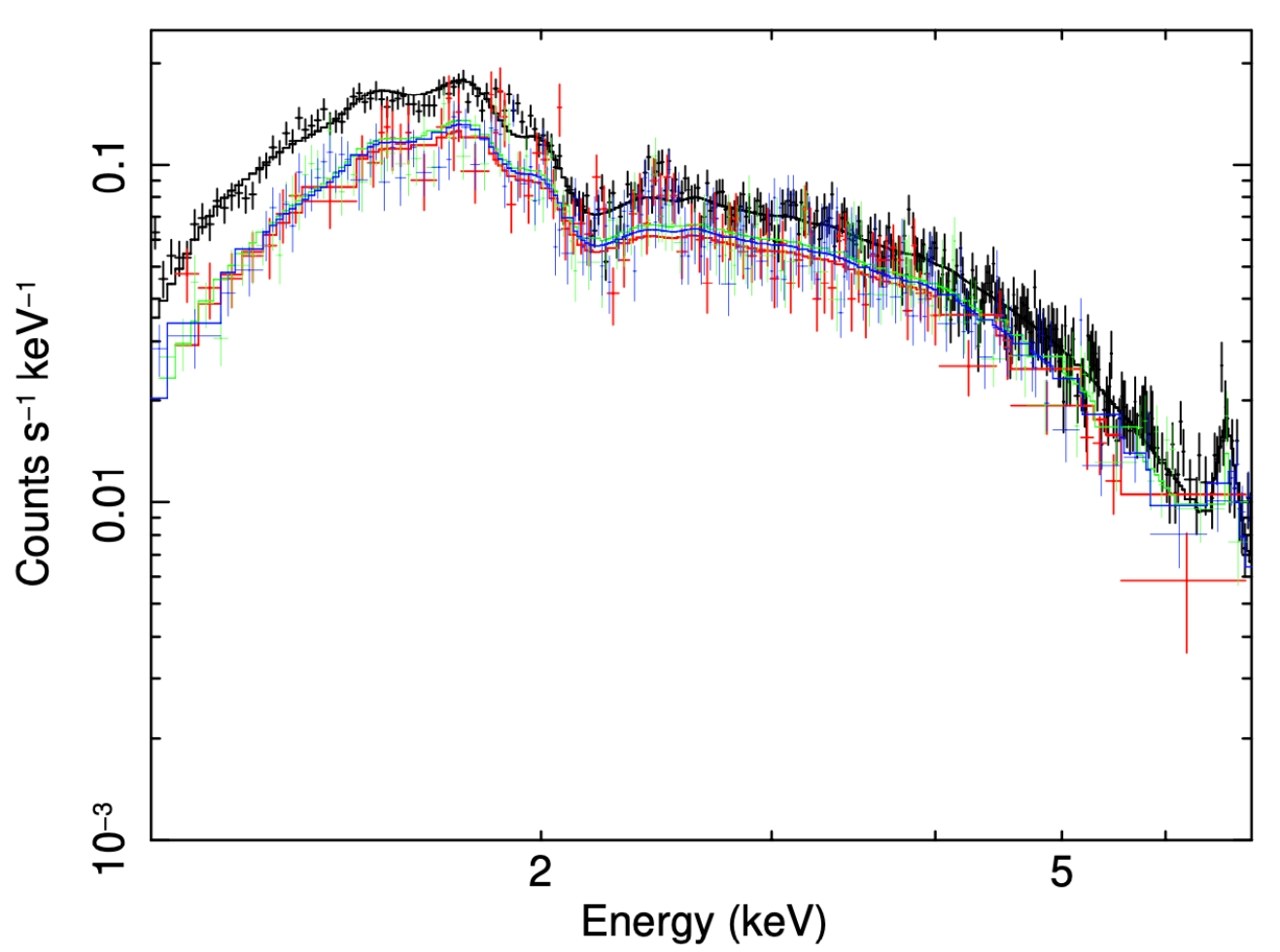}
       \caption{Chandra X-ray spectra of RCW~38 with the best-fit model (which included a thermal and a power-law component). The red points are the data points with error bars and the solid curves represent the best fit models. From the thermal component, we derive the properties of the hot gas, including hot gas temperature $kT$ and electron number density $n_{\rm e}$.} 
       \label{fig:Xrayspectra}
    \end{center}
     \end{figure}

The best-fit X-ray spectra are shown in Figure~\ref{fig:Xrayspectra} which yielded $\chi^2$ = 1196 with 1003 degrees of freedom (a reduced $\chi^2$ = 1.19). The best-fit parameters were the following: a hydrogen column density of $N_{\rm H} = (1.96^{+0.07}_{-0.06})\times10^{22}$~cm$^{-2}$, a hot gas temperature of $kT = 4.52^{+1.22}_{-0.87}$~keV, and an X-ray power-law index of $\Gamma_{\rm X} = 2.04^{+0.22}_{0.17}$.  Nearly half ($\approx$47\%) of the emitted (unabsorbed) flux in the 0.5 $-$ 7.0~keV band is produced by the thermal component (bremsstrahlung and line emission), with $F_{\rm X} = (4.4\pm0.5)\times10^{-12}$~erg~cm$^{-2}$~s$^{-1}$, corresponding to an X-ray luminosity of $L_{\rm X} = (1.5\pm0.2)\times10^{33}$~erg~s$^{-1}$.

To calculate the hot gas electron number density $n_{\rm e}$, we used the best-fit normalization of the thermal component, norm = 3.25$\times10^{-3}$~cm$^{-5}$, which is defined as norm = $(10^{-14}{\rm EM})/4 \pi d^{2}$, where EM is the emission measure, EM = $\int n_{\rm e} n_{\rm H} dV$, and $d$ is the distance to RCW~38. Relating $n_{\rm e}$ to the hydrogen number density by $n_{\rm e}$ = 1.2 $n_{\rm H}$ (the relation for a fully-ionized plasma that is primarily comprised of hydrogen and helium; \citealt{Sarazin1986}), then $n_{\rm e} = (1.5\times10^{15}$~norm~$d^{2}/fV)^{1/2}$, where $f$ is the filling factor of the hot gas and $V$ is the hot gas volume. Assuming a spherical volume with radius $R = 2$\arcmin $\approx$ 1~pc and $f = 1$, we find $n_{\rm e} = 1.9$~cm$^{-3}$. Assuming fully ionized hydrogen, the thermal pressure from the hot gas is given by $P_{\rm X} = 2 n_{\rm e} kT =  2.7\times10^{-8}$~dyn~cm$^{-2}$, which we then compare to our estimated CR pressure $P_{\rm CR}$ in Section~\ref{sec:CR_pressure}.

\section{Discussion} \label{sec:discussion}

In Section~\ref{sec:fermianalysis}, we show that RCW~38 produces substantial $\gamma$-ray emission that is spatially coincident with its young star cluster. This emission may be attributed to CRs (protons and/or electrons) accelerated by the stellar winds. Due to RCW 38’s young age, no SNe have exploded in the region, making it unlikely that the $\gamma$-ray emission is associated with CRs accelerated by SNe. In this section, we aim to constrain the CR acceleration efficiency of the winds and the diffusion coefficient $D$ in the star-forming region. We start by calculating the wind power of the star cluster associated with RCW~38 in Section~\ref{sec:wind_power}, and then in Section~\ref{sec:CRinjection}, we show that in order to produce the detected $\gamma$-ray emission, the acceleration efficiency must be high and/or the diffusion coefficient $D$ is small relative to value typically assumed for the nearby ISM \citep{Strong10}. In Section~\ref{sec:CR_pressure}, we evaluate the CR pressure and compare it to the thermal pressure $P_{\rm X}$ derived in Section~\ref{sec:Xrayanalysis}. Finally, in Section~\ref{sec:leptons}, we demonstrate that the $\gamma$-ray emission in the region is likely due to hadronic CR losses and the observed emission is unlikely to originate from the interaction of accelerated electrons with the ambient medium.

\subsection{Wind Power}
\label{sec:wind_power}

Due to their high luminosities and large escape velocities, massive stars launch fast ($v_{\rm w} \gtrsim 10^3~\rm{km/s}$), line-driven stellar winds \citep[e.g., see reviews by ][]{Smith2014a, Vink2022a}. The collective kinetic energy injection rate per massive star in a YMC is $L_{\rm w} = (1/2) \dot{M}_{\rm w} v_{\rm w}^2$, where $\dot{M}_{\rm w}$ is the stellar mass-loss rate that depends on the stellar properties, such as the star's bolometric luminosity ($L_{\rm bol}$) and surface temperature ($T_{\rm eff}$; \citealt{Vink2001}).

To estimate the total mechanical wind luminosity of RCW~38 (i.e., $L_{\rm w} = (1/2) \sum^{N_{\rm OB}}_{i}\dot{M}_{\rm w} v_{\rm w,~i}^2$) we apply the empirical relationship from \citet{Howarth}
\begin{equation}
    \log_{10}\left(\dot{M}_{\rm w}/M_\odot \ \rm yr^{-1}\right) = 1.69 \ \log_{10}\left(L_{\rm bol}/L_\odot\right) - 15.4,
\end{equation}
\noindent
to the observed $L_{\rm bol}$ OB candidates from \citet{Wolk06} (31 sources). We find a total YMC mass-loss rate of $3.8 \times 10^{-6}~{M_{\odot}\rm~yr^{-1}}$ corresponding to the total $L_{\rm bol} \rm \ of \ 3.7 \times 10^{39}$ erg~s$^{-1}$. 

Since the radii and masses of these sources are not constrained, we are unable to determine the expected escape speeds of the sources to estimate $v_{\rm w,~i}$ following  \citet{Vink2001} who employ the estimate $v_{\rm w} \propto v_{\rm esc}$. Instead, we estimate $v_{\rm w}$ for each OB candidate identified by \citet{Wolk06} in RCW~38 by applying the wind-luminosity relationship given by $0.5 L_{\rm bol}/c = \dot{M}_{\rm w} v_{\rm w}$, which assumes that the momentum flux carried by stellar winds ($\dot{M}_{\rm w}v_{\rm w}$) is approximately half of the radiative momentum flux \citep[$1/2~L_{\rm bol}/c$; ][]{Lopez2011,lancaster21}. This assumption is true for YMCs, where $L_{\rm bol}/c$ ranges from $(0.5 - 3) \dot{M_w} v_{\rm w}$. With these inferred $v_{\rm w,~i}$, we estimate the total mass-loss weighted cluster wind velocity as
\begin{equation}
\langle v_w \rangle_{\dot{M}_{\rm w}} = \frac{\sum^{N_{\rm OB}}_i \dot{M}_{w,~i} v_{w,~i}}{\sum^{N_{\rm OB}}_i \dot{M}_{w,~i}}, 
\end{equation}
and obtain $\langle v_w \rangle_{\dot{M}_{\rm w}} = 2.6\times 10^3~\rm km \ s^{-1}$. These calculations yield a total wind luminosity of $L_{\rm w} = 8 \times 10^{36}~\rm erg \ s^{-1}$ in agreement with the estimated value, $L_{\rm w} \gtrsim 3 \times 10^{36}~\rm erg \ s^{-1}$, presented in \citet{Ge2024}. 

Our estimated value for $L_{\rm w}$ is roughly an order of magnitude less than that of \citet{Peron24} who assumed $\dot{M}_{w} = 10^{-4}~M_{\odot}~yr^{-1}$ and $v_w = 1000~\rm{km~s^{-1}}$, yielding $L_{\rm w}\sim$$6 \times 10^{37}~\rm{erg~s^{-1}}$, following the fiducial values presented in \citet{Canto00}. As a check our estimated value, we also compute this quantity from {\fontfamily{cmtt}\selectfont STARBURST99} \citep{Leitherer1999}. We note that since RCW~38 has a total stellar mass of $\sim2.5 \times 10^3~M_{\rm \odot}$, the IMF is most likely sensitive to stochastic sampling rather than being fully-sampled as expected for clusters with $M_{\rm \star} \lesssim 10^5~{M_{\rm \odot}}$ \citep{DaSilva2012}. With these caveats, we adopt the observed IMF slopes from \cite{Muzic}, the total stellar mass of the cluster as $M_{\rm tot} = 2251 \  M_\odot$, and $40~ M_\odot$ as the maximum mass of a star (see Table B1 of \citealt{Weidner10}). Assuming solar metallicity, we get $L_{\rm w}$ = $6.79 \times 10^{36}$ erg s$^{-1}$, which is comparable to our previous estimate for $L_{\rm w}$ derived using the wind momentum-luminosity relation value.

\subsection{Cosmic-ray Injection}
\label{sec:CRinjection}

With $L_{\rm w}$ derived above, we now investigate the acceleration efficiency of CRs from stellar winds in RCW~38. The energy injection rate into CRs by stellar winds is $\dot{E}_{\rm CR} = \eta_{\rm CR} L_{\rm w}$, where $\eta_{\rm CR}$ is the fraction of the wind kinetic energy that goes into accelerating primary CR protons. 

Given $\dot{E}_{\rm CR}$, the maximum $\gamma$-ray luminosity $L_{\gamma}^{\rm max}$ that can be produced by the CRs in inelastic proton-proton collisions is $L_{\gamma}^{\rm max} = f_\gamma \dot{E}_{\rm CR}$, where $f_\gamma = 1/3$ represents the fraction of the energy that goes into neutral pions, which then decay to $\gamma$-rays. We define the calorimetry fraction $f_{\rm cal}$ as the ratio of the observed $\gamma$-ray luminosity $L_{\gamma}$ to the maximum $\gamma$-ray luminosity $L_\gamma^{\rm max}$: $f_{\rm cal} \equiv L_\gamma / L_\gamma^{\rm max}$. A low inferred value of $f_{\rm cal}$ can be attributed to CR escape via diffusion from the star cluster. Putting these terms together, we have 
\begin{equation}
f_{\rm cal} =  \frac{3 L_{\gamma}}{\eta_{\rm CR}L_{\rm w}}.
\label{eq:fcal}
\end{equation} 
\noindent
If the $\gamma$-rays are produced via pionic emission when CR protons collide with gas, and if the escape losses are dominated by diffusion, then $f_{\rm cal}$ can also constrain the ratio of those two timescales.  The CR diffusion timescale $t_{\rm diff}\sim R^2/D$ is of order
\begin{equation}
t_{\rm diff} = 1500~{\rm yr} \bigg( \frac{R}{7~{\rm pc}} \bigg)^{2} \bigg( \frac{10^{28}~{\rm cm}^2~{\rm s}^{-1}}{D}\bigg),
\label{eq:tdiff}
\end{equation} 
where $R$ is the radius of the spherical region where diffusion takes place (estimated below) and $D$ is the diffusion coefficient normalized to a typical value in the galactic ISM \citep{Galprop}. 
From our $\gamma$-ray extension analysis in Section~\ref{sec:extension}, RCW~38 has $\theta\approx0.23^{\circ}$, corresponding to a physical radius of $R = 6.8$~pc for a distance of $d = 1.7$~kpc.

The pion loss timescale $t_\pi$, which corresponds to the timescale for proton-proton hadronic interaction losses, is \citep{Mannheim}
\begin{equation}
t_\pi = 5\times10^{4}~{\rm yr} \bigg ( \frac{n_{\rm eff}}{10^{3}~{\rm cm}^{-3}}\bigg)^{-1},
\label{eq:t_pi}
\end{equation}
where $n_{\rm eff}$ is the effective number density of nucleons encountered by CRs (see $n_{\rm eff}$ estimate below). If $t_\pi\ll t_{\rm diff}$ and if there are no other losses (e.g., CR streaming losses), then $f_{\rm cal}\rightarrow1$. Conversely, if $ t_{\rm diff} < t_{\pi}$ then $f_{\rm cal}\simeq t_{\rm diff}/t_\pi$. 

Using these relations and assumptions, our estimates of $L_{\rm w}$ from Section \ref{sec:wind_power}, and our determinations of $L_\gamma$ and $R$ from the Fermi observations (note that the $\gamma$-ray extent is larger than the X-ray extent from Section~\ref{sec:Xrayanalysis} since the CR protons need to encounter dense gas before pions are produced), we can constrain $\eta_{\rm CR}$ and $D$ given $n_{\rm eff}$. In particular, assuming that $f_{\rm cal}=t_{\rm diff}/t_{\pi}<1$, Equation~\ref{eq:fcal} can be rewritten in terms of these observables: 
\begin{eqnarray}
\eta_{\rm CR} \bigg( \frac{10^{28}~{\rm cm}^2~{\rm s}^{-1}}{D}\bigg) \bigg( \frac{n_{\rm eff}}{10^{3}~{\rm cm}^{-3}} \bigg) =  \nonumber \\
\frac{5\times10^{4}}{1500} \bigg( \frac{3 L_{\gamma}}{L_{\rm w}} \bigg) \bigg( \frac{7~{\rm pc}}{R} \bigg)^{2}.
\label{eq:constraint}
\end{eqnarray}
Using $L_\gamma = 2.6\times10^{34}$~erg s$^{-1}$ (Section~\ref{sec:spectral_analysis}) and $L_{\rm w} = 8\times10^{36}$~erg s$^{-1}$ (Section~\ref{sec:wind_power}),  $3L_\gamma/L_{\rm w} \simeq 0.01$. 

Finally, we estimate $n_{\rm eff}$. \cite{Fukui16} report a total gas mass in the two colliding molecular clouds associated with RCW~38 of $2.3 \times 10^{4}~M_\odot$. Assuming a spherical volume with radius $R = 6.8$~pc, we find an effective number density of $n_{\rm eff}= 780$~cm$^{-3}$. 

As an independent check, we can constrain the local gas density from our X-ray analysis. The hydrogen column density $N_{\rm H}$ inferred from the spectral analysis in Section~\ref{sec:Xrayanalysis} is $N_{\rm H} = (1.96^{+0.07}_{-0.06})\times10^{22}$~cm$^{-2}$. If this absorbing column is due to gas in close proximity to RCW~38, then $n_{\rm eff} = N_{\rm H} / R \approx 930$~cm$^{-3}$. Both estimates of $n_{\rm eff}$ are similar order-of-magnitude.

Adopting $n_{\rm eff} \sim 10^{3}$~cm$^{-3}$, then 
\begin{equation}
\frac{t_\pi}{t_{\rm diff}}\simeq30\left(\frac{10^3\, {\rm cm^{-3}}}{n_{\rm eff}}\right)\left(\frac{7\,{\rm pc}}{R}\right)^2\left(\frac{D}{10^{28}\,{\rm cm^2\,\,s^{-1}}}\right),
\label{tpi_tdiff}
\end{equation}
suggesting that diffusion losses dominate over pion losses for our estimated values and that Equation~\ref{eq:constraint} holds.

Rewriting Equation~\ref{eq:constraint} to solve for $\eta_{\rm CR}$, we have
\begin{eqnarray}
    \eta_{\rm CR}\,&\simeq&\,0.4\left(\frac{10^3\, {\rm cm^{-3}}}{n_{\rm eff}}\right)\left(\frac{7\,{\rm pc}}{R}\right)^2 \nonumber \\
    &\,&\times \left(\frac{D}{10^{28}\,{\rm cm^2\,\,s^{-1}}}\right)\left(\frac{3L_\gamma/ L_{\rm w}}{0.01}\right),
\label{eq:high_eta}
\end{eqnarray}
implying a high acceleration efficiency for stellar winds of $\eta_{\rm CR} \sim 0.4$ with a diffusion coefficient typical of the nearby ISM.

Alternatively, assuming the value of $\eta_{\rm CR}=0.1$ which is typical of SN shocks \citep{Vink12}, this expression can be written as a constraint on the diffusion coefficient $D$:
\begin{eqnarray}
    D\,&\simeq&\,2.5\times10^{27}\,{\rm cm^2\,\,s^{-1}} \left(\frac{\eta_{\rm CR}}{0.1}\right)
    \left(\frac{n_{\rm eff}}{10^3\, {\rm cm^{-3}}}\right) \nonumber \\
    &\,&\times\left(\frac{R}{7\,{\rm pc}}\right)^2 
    \left(\frac{0.01}{3L_\gamma/L_{\rm w}}\right).
    \label{bigD}
\end{eqnarray}

\noindent
The above scaling demonstrates that this lower CR acceleration efficiency of $\eta_{\rm CR} = 0.1$ necessitates a smaller diffusion coefficient of $D \sim 2.5\times10^{27}$~cm$^{2}$~s$^{-1}$ in the vicinity of the star cluster. 

These results conform with our intuition that more rapid diffusion (escape), via either bigger $D$ or smaller $R$, requires a higher CR efficiency $\eta_{\rm CR}$ to maintain the same observed $\gamma$-ray emission relative to the power provided by winds. Similarly, larger $n_{\rm eff}$ trades off against escape losses via the ratio $t_\pi/t_{\rm diff}$. Note that for the value of $D$ in Equation~\ref{bigD}, $t_\pi/t_{\rm diff}\simeq8$ in Equation~\ref{tpi_tdiff}, indicating that losses are only marginally dominated by diffusion for the adopted gas density and that $f_{\rm cal}\simeq0.1$. 

Overall, we see that a hadronic interpretation of the observed $\gamma$-ray emission passes basic checks: for $\eta_{\rm CR}$ in the range of $\simeq0.1$ and $D\simeq2.5\times10^{27}$\,cm$^2$ s$^{-1}$, the $\gamma$-ray luminosity is reproduced if $n_{\rm eff}$ is of the order $10^3$\,cm$^{-3}$. However, if the CRs instead interact with a medium well below this nominal gas number density, the diffusion coefficient $D$ must be proportionately smaller to maintain the same $L_\gamma/L_{\rm w}$, as made explicit in Equations~\ref{bigD} and \ref{tpi_tdiff}. We note that our constrained value of $D$ is in accordance with theoretical estimates provided in \cite{Gupta1}.

\subsection{CR pressure}
\label{sec:CR_pressure}

With these numbers and scalings in hand, the CR pressure $P_{\rm CR}$ in the region can be roughly estimated from 
\begin{equation}
    P_{\rm CR}\simeq \frac{1}{3}\frac{\dot{E}_{\rm CR}\min[t_\pi,t_{\rm diff}]}{V} ,
\end{equation}
where $V$ is the region's volume. Assuming again that $t_{\rm diff}<t_\pi$ (see Equation~\ref{tpi_tdiff}), we have that 
\begin{eqnarray}
    P_{\rm CR}&\simeq&\frac{\eta_{\rm CR} L_{\rm w}}{4\pi RD}\simeq 1\times10^{-12}\,{\rm erg\,\,cm^{-3}}\left(\frac{\eta_{\rm CR}}{0.1}\right)\nonumber \\
    &\times&\left(\frac{L_{\rm w}}{10^{37}\,{\rm erg\,\,s^{-1}}}\right)\left(\frac{7\,{\rm pc}}{R}\right)\left(\frac{10^{27}\,{\rm cm^2\,\,s^{-1}}}{D}\right).\,\,\,\,\,\,
\label{eq:P_cr}
\end{eqnarray}
We can also use Equation~\ref{eq:constraint} to write $P_{\rm CR}$ in terms of quantities connected with observations: $P_{\rm CR}\simeq 3L_\gamma t_\pi/4\pi R^3$:
\begin{eqnarray}
    P_{\rm CR}&\simeq&1\times10^{-12}\,{\rm erg\,\,cm^{-3}}\left(\frac{L_\gamma}{2.6\times10^{34}\,{\rm ergs\,\,s^{-1}}}\right) \nonumber \\
    &\times&\left(\frac{7\,{\rm pc}}{R}\right)^3\left(\frac{10^3\,{\rm cm^{-3}}}{n_{\rm eff}}\right).\,\,\,\,\,\,\,\,\,\,
\end{eqnarray}
These estimates are equivalent to the method of \cite{Gupta1}.

In Section~\ref{sec:Xrayanalysis}, we measured the thermal pressure of the hot gas to be $P_{\rm X} = 2.7\times10^{-8}$~dyn~cm$^{-2}$, roughly four orders of magnitude greater than $P_{\rm CR}$. We note that the thermal pressure is calculated within a region roughly one pc in size, whereas the $P_{\rm CR}$ is calculated for a region seven times larger. This difference arises because the X-rays are confined to a smaller volume characterized by a decreasing density and a declining thermal pressure, as shown in Figure~1 of \cite{Gupta2}. Our results suggest that CR feedback is most likely dynamically unimportant in very young ($t \lesssim 0.5$~Myr) YMCs. However, we note that the inferred $P_{\rm X}$ in RCW~38 is much larger than those estimated in $\sim$1-3~Myr YMCs \citep[e.g., see ][]{Rosen2014a,Gupta1}. Thus, it still remains uncertain if CR feedback may be dynamically important in slightly older YMCs that contain cooler (e.g., $\sim~1-5 \times 10^6$~K) and more extended superbubbles.

\subsection{Estimating the Leptonic $\gamma$-ray Emission}
\label{sec:leptons}

In order to estimate if the observed $\gamma$-ray emission can be leptonic in origin, we evaluate the $L_\gamma$ assuming a purely leptonic model. $\gamma$-ray emission can be produced via relativistic bremsstrahlung and inverse-Compton scattering of the primary CR electrons. In the latter scenario, low-energy photons such as CMB, optical, and UV (stellar radiation) can be inverse-Compton boosted by relativistic electrons to $\gamma$-ray energies. The energy of the scattered photons is approximately $ E_{\rm IC}  \sim \Gamma^2 \ E_{\rm seed}$ where $\Gamma$ is the Lorentz factor of the relativistic electron and $E_{\rm seed}$ is the energy of the seed photon (usually ranging from $0.1-10$ eV for stellar radiation). For an optical photon of 1\,eV, the Lorentz factor of the relativistic electron should be of the order of $\sim 10^{4.5}$ such that the IC boosted $\gamma$-ray photon has the energy of $\sim 1$ GeV. The timescale for inverse-Compton emission is given by
\begin{eqnarray}
    t^{\rm e}_{\rm IC}&=&\rm \frac{\Gamma m_{\rm e} c^2}{ \frac{4}{3} \sigma_{\rm T} c \Gamma^2~U_{\rm ph}}\simeq 3\times10^6\,\rm yr \left(\frac{10^{4.5}}{\Gamma}\right)\nonumber \\
    &\times&\left(\frac{10^{-11}\rm erg\,\,cm^{-3}}{\rm U_{ph}}\right),
    \label{eq:t_IC}
\end{eqnarray}
where $m_{\rm e}$ is the rest mass of electron, $\sigma_{\rm T}$ is the Thomson cross-section, $c$ is the speed of light, and $U_{\rm ph}$ is the stellar radiation energy density computed as $\rm U_{ph} \simeq  (L_{\rm bol}/4\pi r^2 c)$. We use the values $L_{\rm bol} =3.7 \times 10^{39}$~erg~s$^{-1}$ (Section \ref{sec:wind_power}) and $R\approx 7$ pc to determine $U_{\rm ph}$. Note that $t_{\rm IC}$ is much longer than the diffusion timescale for a broad range of $D$ (eq.~\ref{eq:tdiff}).

Next, we estimate the $\gamma$-ray luminosity from IC emission. We assume that in the star-forming region, the dominant radiation field is the stellar radiation. For a given population of relativistic electrons, the number density distribution can be defined as $n(\Gamma) = \kappa_1 \Gamma^{-p}$, where $p$ is the spectral index of relativistic electrons ($p \approx$ 2.2), and $\kappa_1$ is the normalization constant that can be calculated from the energy density of CR electrons ($\epsilon_{CR\_e}$) as follows \citep{Gupta2}:
\begin{equation}
    \kappa_1 \approx \rm \frac{\epsilon_{CR\_e}}{m_e c^2} (p-2) \left[\frac{1}{\Gamma^{p-2}_{min}} - \frac{1}{\Gamma^{p-2}_{max}}\right ]^{-1}.
\end{equation}
\noindent 
The upper-limit on the inverse-Compton $\gamma$-ray luminosity ($L_{\gamma}^{\rm IC}$) can be determined by \citep{Gupta2}:
\begin{equation}
    L_{\gamma}^{\rm IC} =  \int_{V} dV \left[ \rm \frac{4}{3} \ \sigma_T \ c \ U_{ph}\ \kappa_1 \ \frac{\Gamma_{\rm max}^{(3-p)}- \Gamma_{\rm min}^{(3-p)}}{3-p}\right],
\end{equation}
where $V$ is the volume. We take $\Gamma_{\rm min}=1$ and $\Gamma_{\rm max}=10^6$. Based on the calculations above, and assuming  $L_{\gamma}^{\rm IC} \simeq L_{\gamma}^{\rm obs}$, we find that the implied value of the CR electron energy density $\epsilon_{\rm CR_e}$ must be 1000 times higher than the CR proton energy density under the assumptions of Section \ref{sec:CR_pressure}. This in turn implies that for primary CR electrons, the acceleration efficiency would need to be $\eta_{\rm CR_e} \gg 1$ (from Eqn.~\ref{eq:P_cr}). Thus, a leptonic model where the observed $\gamma$-ray emission is dominated by inverse-Compton emission from primary CR electrons interacting with starlight from the cluster is strongly disfavored. 

Importantly, our effective number density ($n_{\rm eff}$) estimates in Section \ref{sec:CRinjection}, derived from X-ray gas density and CO observations, indicate the presence of a dense medium. If primary CR electrons are interacting with such a dense medium then relativistic bremsstrahlung will be dominant over inverse-Compton losses. To see this, note that the timescale for relativistic bremsstrahlung losses is given by \citep{BlumenthalGould1970}
\begin{equation}
    t^e_{\rm brems} \simeq 4 \times 10^4~ \text{yr}~ \left(\frac{10^3\,{\text{cm}^ {-3}}}{n_{\rm eff}}\right).
\end{equation}
Note that we have used $\Gamma \simeq 10^{4.5}$ to estimate the cooling timescale above, but the energy dependence of $t_{\rm brems}^{\rm e}$ is weak. The relativistic bremsstrahlung loss timescale is similar to $t_{\pi}$ (Eq.~\ref{eq:t_pi}), with the same dependence on $n_{\rm eff}$.  The ratio of the relativistic bremsstrahlung and IC cooling timescales is $\simeq 0.01$ (Eq. \ref{eq:t_IC}), suggesting that relativistic bremsstrahlung could dominate inverse-Compton emission if CR electrons interact with the average density medium. 

However, if CR electrons interact with the average density medium, CR protons would too. Since we expect the CR acceleration efficiency for protons to be larger than for primary electrons $\eta_{\rm CR_p} \gg \eta_{\rm CR_e}$, we would thus expect pion losses from CR proton collisions with the ambient medium to dominate the observed $\gamma$-ray emission. Therefore, a hadronic origin for the $\gamma$-rays emission appears more plausible than a solely leptonic origin via either inverse-Compton or relativistic bremsstrahlung.

\subsection{Comparison to Previous Fermi Work on RCW~38}
\label{sec:Comparison}

During the completion of this manuscript, work by \cite{Peron24} was published which provides an independent analysis of Fermi data toward several young star clusters in the Vela molecular ridge, including RCW 38. They reported $L_\gamma (> 1 \rm GeV) = 5 \times 10^{33}$~erg~s$^{-1}$ based on a power-law model with an index of $\Gamma = 2.56\pm0.05$ at an assumed distance of $d = 1.6$ kpc. If placed at our assumed distance of $d = 1.7$~kpc, their luminosity becomes $L_\gamma (>1~{\rm GeV}) = 5.6\times10^{33}$~erg~s$^{-1}$, which is comparable to our calculated value of $L_{\gamma}(>1~{\rm GeV}) = 5.2\times10^{33}~{\rm erg}~{\rm s}^{-1}$ (Section \ref{sec:spectral_analysis}). 

Additionally, \cite{Peron24} estimated an upper limit on the wind power from the star cluster using the \cite{Weaver} model that relates the observed bubble radius to wind power, ambient density, and age. They assumed a mass-loss rate of $\dot{M} = 2\times 10^{-4}~{M_{\odot}\rm~yr^{-1}}$, which is higher than our estimated value $\dot{M} = 3.8 \times 10^{-6}~{M_{\odot}\rm~yr^{-1}}$ (see Sec.~\ref{sec:wind_power}). They further calculated $\dot{E}_{\rm CR}$ using a calorimetric assumption (i.e., all accelerated CRs lose their energy via pion decay), finding a lower limit on $\eta_{\rm CR}$ for RCW~38 to be $\approx 0.004\%$ (see their Table~1). If we adopt their assumed mass-loss rate and only consider $L_{\gamma}$ above 1 GeV, then we would find $\eta_{\rm CR}$ becomes $\approx 0.003\%$, similar to their value. Thus, the discrepancy in the derived $\eta_{\rm CR}$ values can be mostly attributed to the difference in wind luminosity estimates: their reported wind luminosity is $\simeq 6 \times 10^{37}$~erg~s$^{-1}$, which is greater than our value of $8 \times 10^{36}$~erg~s$^{-1}$ (Sec.~\ref{sec:wind_power}), leading to a much lower value of $\eta_{\rm CR}$.

\cite{Ge2024} also reported the Fermi-LAT detection of RCW 38 using data above 1 GeV energy range. They reported $L_\gamma (> 1 \rm GeV) = 1.6 \times 10^{33}$~erg~s$^{-1}$ based on a power-law model with an index of $\Gamma = 2.44\pm0.03$ at an assumed distance of $d = 1.7$ kpc. In addition, they estimated an effective gas density in the region to be $328 \ \rm cm^{-3}$ using the CO composite survey. This value is lower as compared to our calculated value of $ \rm n_{eff} = 1000 \ cm^{-3}$. Using standard literature values of $\dot{M}$ and $v_{\rm w}$, they constrain the wind power to be $3 \times 10^{36}$~erg~s$^{-1}$, which is smaller than our value of $8 \times 10^{36}$~erg~s$^{-1}$. They assumed a CR acceleration efficiency of 10 percent. They derived a diffusion coefficient value of $4\times10^{26}\,{\rm cm^2\,\,s^{-1}}$, which is an order of magnitude lower than our estimated value of $2.5\times10^{27}\,{\rm cm^2\,\,s^{-1}}$ (see Eqn.\ref{bigD}), for the same CR acceleration efficiency. The discrepancy in the values can be associated with their lower estimate of the total number density of gas in the region. 

Overall, the three papers agree that stellar wind feedback from YMCs produce CR protons that are detectable by Fermi.

\section{Conclusions} 
\label{sec:conclusions}

We report the significant $\gamma$-ray detection ($\sigma > 22$) of the young ($<$0.5~Myr old) star cluster in the star-forming region RCW~38 using 15 years of Fermi-LAT data. We find that the Fermi source 4FGL J0859.2$-$4729 is potentially associated with the star cluster and has an angular extent of $\approx 0.23\pm0.04^\circ$ (Section \ref{sec:likelihood} and Section \ref{sec:extension}). The $\gamma$-ray SED follows a PL distribution with a photon index of $\Gamma = 2.34\pm0.04$ (Figure \ref{fig:sed}), and we estimate the luminosity to be $L_\gamma$ =  $(2.66\pm0.92)\times10^{34}~{\rm erg}~{\rm s}^{-1}$ in the 0.1 $-$ 500~GeV band (Section \ref{sec:spectral_analysis}). 
 
Our estimates suggest that the observed $\gamma$-ray emission is hadronic in origin, resulting from pion production as CR protons interact with the high-density ambient medium. Purely leptonic models via inverse-Compton or relativistic bremsstrahlung are disfavored under simple assumptions about the primary CR acceleration efficiency (Section \ref{sec:leptons}). Given that no SNe have occurred in a star cluster of this age, the detected $\gamma$-ray emission likely arises from CRs accelerated by stellar winds. 

We use the observed bolometric luminosity of the OB candidates in the star cluster to determine the total stellar wind luminosity of $L_{\rm w}=8\times10^{36}$ erg s$^{-1}$ (Section \ref{sec:wind_power}). With this estimate, we constrain the CR acceleration efficiency $\eta_{\rm CR}$ and the diffusion coefficient $D$ together near the star cluster. We demonstrate that the observed $\gamma$-ray luminosity necessitates a high CR efficiency of $\eta_{\rm CR} \simeq$ 0.4 if the diffusion coefficient is consistent with the local ISM value of $D = 10^{28}$~cm$^{2}$~s$^{-1}$. Alternatively, a CR efficiency of $\eta_{\rm CR}\simeq0.1$ is possible if the diffusion coefficient is smaller around the region, $D\simeq 2.5\times10^{27}\,{\rm cm^2\,\,s^{-1}}$ (Section \ref{sec:CRinjection}; eqs.~\ref{eq:high_eta}, \ref{bigD}). 

We estimate a CR pressure in the region of $P_{\rm CR} = 1\times10^{-12}\ \rm erg \ cm^{-3}$ (Section \ref{sec:CR_pressure}), roughly four orders of magnitude lower than the thermal pressure of the hot gas $P_{\rm X}= 2.7\times10^{-8}$~dyn~cm$^{-2}$ (Section \ref{sec:Xrayanalysis}). Our results suggest that CR feedback is dynamically less important than the thermal pressure from stellar wind feedback in RCW~38's young cluster.
 
Our work adds to the growing body of literature establishing that stellar winds from YMCs contribute to the galactic CR population. Similar analyses on other YMCs would be valuable to establish how the CR acceleration, transport, and dynamics evolve with cluster properties and age. 

\software{CIAO (v4.14; \citealt{CIAO2006}), XSPEC (v12.12.1; \citealt{XSPEC}), Fermitools (v11r5p31), FermiPy python package \citep{Wood_fermipy}}.

This paper employs a list of Chandra datasets, obtained by the Chandra X-ray Observatory, contained in the Chandra Data Collection (CDC) ~\dataset[doi:291]{https://doi.org/10.25574/cdc.291}.

\begin{acknowledgements}

PP is grateful to Lachlan Lancaster, Biman Nath, and Jordan Eagle for their valuable input and constructive feedback. LAL acknowledges support through the Heising-Simons Foundation grant 2022-3533. LAL and LL gratefully acknowledge the support of the Simons Foundation. ALR acknowledges support from the National Science Foundation (NSF) Astronomy and Astrophysics Postdoctoral Fellowship under award AST-2202249. TAT and IB are supported in part by NASA grant 80NSSC23K1480. TL is supported by the European Research Council under grant 742104, the Swedish National Space Agency under contract 117/19 and the Swedish Research Council under contract 2022-04283.

\end{acknowledgements}

\bibliography{paper}{}
\bibliographystyle{aasjournal}

\end{document}